\documentclass[%
preprint,
 amsmath,amssymb,
 aps,
]{revtex4-1}
\def\comment#1{}
\usepackage{mathrsfs}
\usepackage{graphicx}
\usepackage{dcolumn}
\usepackage{bm}

\begin{document}

\title{Comment on `` Is a Circular Orbit Possible According to General Relativity?" (Arxiv:1008.3553v1)}

\author{Wen-Biao Han}
 \altaffiliation[Also at ]{Physics Department, University of Rome ``La Sapienza," P.le A. Moro 5, 00185 Rome, Italy}
 \email{wenbiao@icra.it}
\affiliation{%
 ICRANet Piazzale della Repubblica, 10-65122, Pescara, Italy 
}%




\date{\today}
\maketitle

A new paper gr-qc/1008.3553V1 by F. T. Hioe and D. Kuebel published
on 20th Aug present that there is no circular orbit in general
relativity \cite{Hioe}:

``(1) A stable circular orbit is not possible. The most ``circular"
orbit must have a small but nonzero eccentricity and is thus
elliptical and precesses.

... ...

(3) There is no so-called innermost stable circular orbit."

These results are very interesting and very different from the known
theory in popular books and references of general relativity, for
example \cite{Hartle}.

In order to check the results of arxiv:1008.3553v1 \cite{Hioe}, we
do some direct numerical tests by using geodesic equation in general
relativity
\begin{align}
\ddot{x}^{\mu}=-\Gamma^\mu_{\alpha\beta}\dot{x}^\alpha\dot{x}^{\beta}.
\end{align}
There are two constants, one is energy $E$ and the another angular
momentum $L$ when a test particle is along the geodesic in
Schwarzschild metric.

\begin{figure}[!h]
\begin{center}
\includegraphics[width=2.5in]{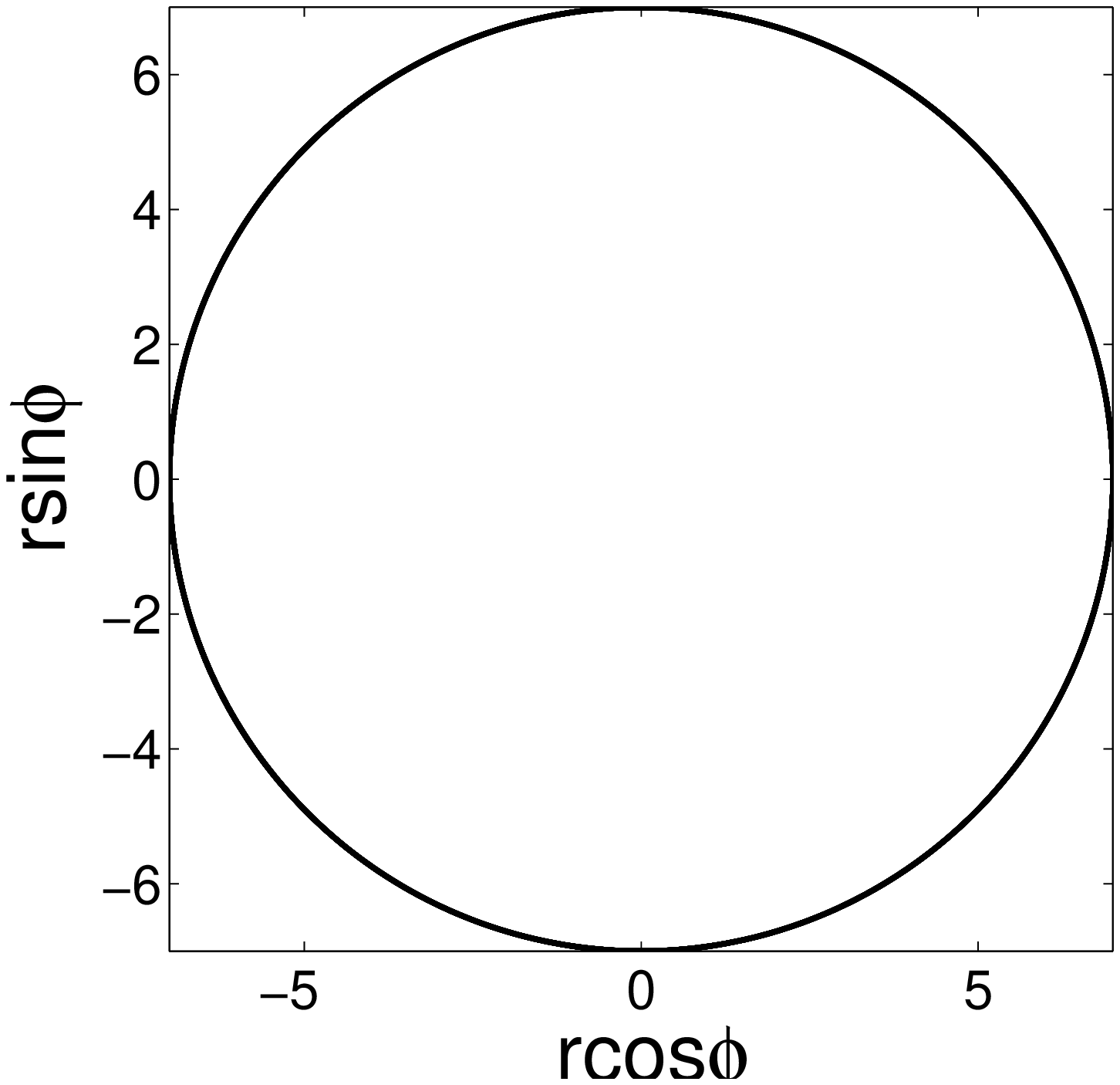}
\includegraphics[width=3.0in]{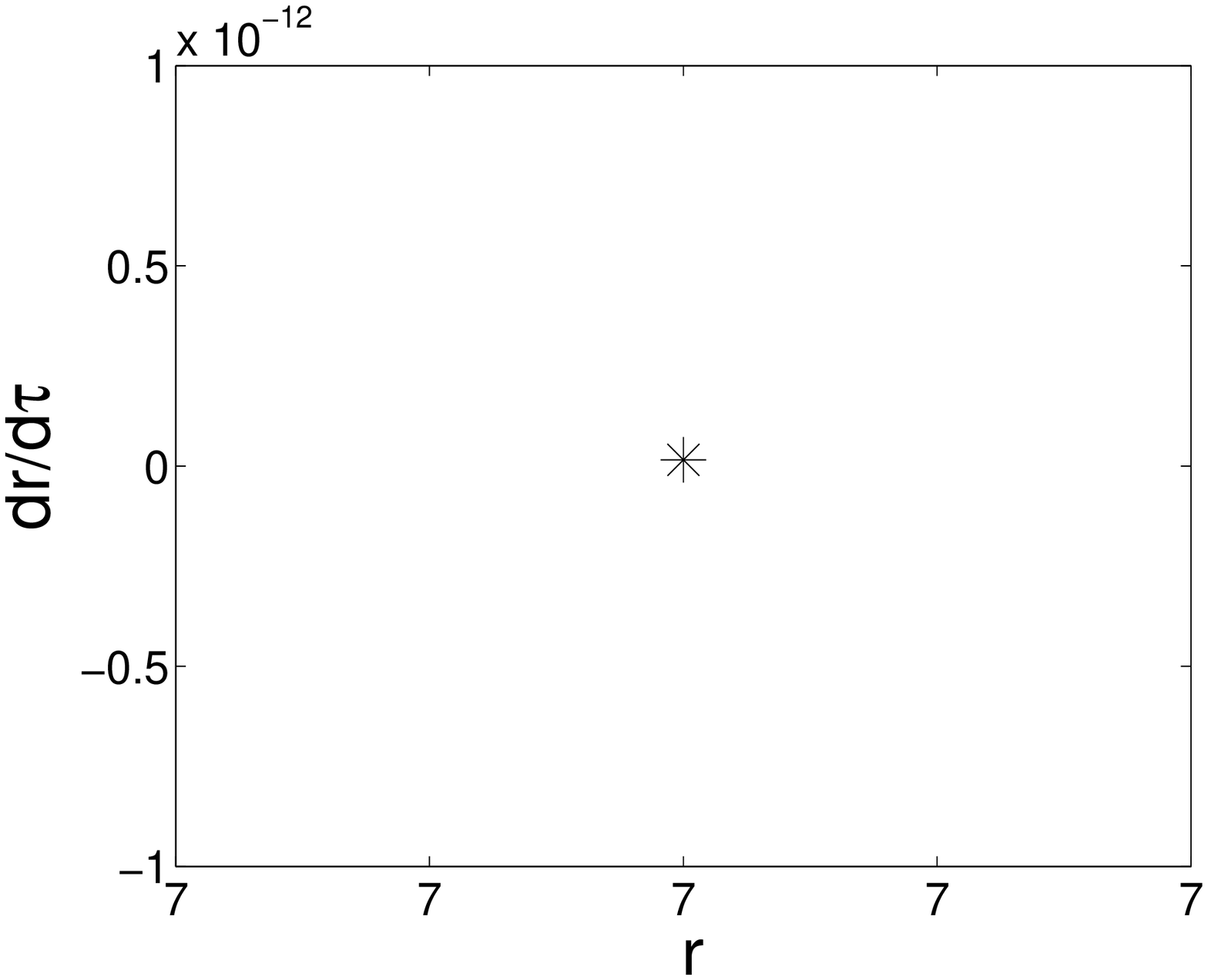}
\caption{Left: The orbit of a test particle with $E=0.944911182523$,
$L=3.5$ and initial radius $r=7$. Right: The Poincar\'e section of
the orbit with same parameters.} \label{r7}
\end{center}
\end{figure}

Giving initial radius $r=7$ and $\theta=\pi/2$, from effective
potential theory, we can decide $E=0.944911182523$ and $L=3.5$ for a
circular orbit. Here we use the unit $G=c=1$ and the mass of black
hole is $1$. Numerical results with a evolution time $t=10^4$ are
shown in Fig~(\ref{r7}). From the Figures, the orbit is circular
with an acceptable error $\sim 10^{-12}$. This error is arisen from
the truncation error of the calculation of $E$.

From Hioe and Kuebel's paper \cite{Hioe}, the orbit with angular
momentum $L=3.5$ must have a true eccentricity
\begin{align}
\epsilon=\sqrt{2}s+\cdots \approx \sqrt{2}GM/hc\approx 0.404,
\end{align}
this does not agree with the above numerical test. And the enough
long evolution time shows that the circular orbit is stable too.

\begin{figure}[!h]
\begin{center}
\includegraphics[width=2.5in]{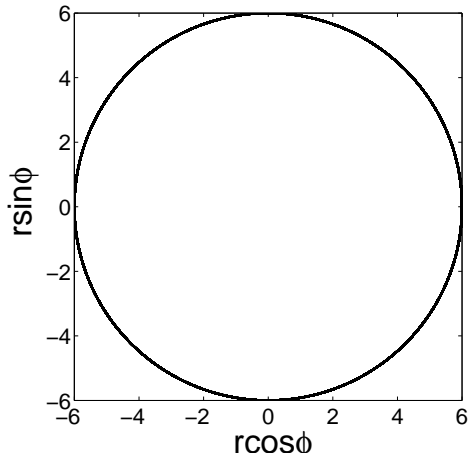}
\caption{The innermost stable circular orbit of a test particle
around Schwarzschild black hole.} \label{isco}
\end{center}
\end{figure}

Finally, we prove the existence of ISCO ($r=6$) in Schwarzschild
spacetime. Setting $E=\sqrt{8/9}$ and $L=\sqrt{12}$, after the same
evolution time, numerical code gives out a perfect stable circular
orbit with zero error (double-precision), see Fig (\ref{isco}).

What induces the disagreement of the analysis of paper \cite{Hioe}
and the numerical experiments? This comment thinks that the problem
is the analysis of Eq.(9) and (10) in \cite{Hioe}. The authors
defined a dimensionless parameter $e$ and required $e \geq 0$. But
actually, we do not require $e\geq 0$ even a real number! Because in
Eq.(9) of \cite{Hioe}, the key variable is $1-e^2$. It has physical
meaning once $1-e^2$ is real. This means that $e$ can be less than 0
or even imaginary number.

We can take the parameters of ISCO into Eq.(10) of \cite{Hioe},
\begin{align}
e=\sqrt{\left[1+\frac{h^2c^2(\kappa^2-1)}{(GM)^2}\right]}=\sqrt{(-1/3)}=i\sqrt{3}/3,
\end{align}
and $1-e^2=4/3$.

In conclusion, we think that the analysis of parameter $e$ in paper
arxiv:1008.3553v1 \cite{Hioe} is incomplete. But the resulting
question is that what is the physical explanation of the energy
eccentricity parameter $e$ defined in \cite{Hioe} when it is an
imaginary number.

\begin{acknowledgments}
Thanks a lot for the discussion with Prof. Hioe.
\end{acknowledgments}

\end{document}